# Generalized thermoelastic band structures of Rayleigh wave in one-dimensional phononic crystals

Ying Wu, Kaiping Yu[*], Linyun Yang, Haotian Zhou and Rui Zhao

*Department of Astronautic Science and Mechanics, Harbin Institute of Technology, Harbin, Heilongjiang 150001, People's Republic of China*

*Corresponding author: yukp@hit.edu.cn (K. Yu).

**Abstract**

We investigate generalized band structures of Rayleigh wave in 1-D phononic crystals in the context of Green–Nagdhi thermoelastic (TE) theory. Solutions of the coupled equations included thermal field are obtained firstly. Then according to boundary conditions, phase velocity is derived. The transfer matrix methodology is adopted in order to formulate the band structures of generalized TE Rayleigh wave. It is demonstrated that the band structure of TE Rayleigh wave is comprised of elastic and thermal bands. For Aluminum/Epoxy phononic crystal, the thermoelasticity can influence the transmission ability of band gaps as well as narrow down the band gap width.

**Introduction**

From the 1960s, the generalized thermoelasticity problem attracts researchers and becomes increasingly significant due to its extensive applications in diverse fields. In the nuclear field, the designations and operations are affected by the extremely high temperatures and temperature gradients originating inside nuclear reactors [1]. Besides, wave motion based on thermoelasticity plays a great role in porous ceramics and reticulate porous that can be used as lightweight structural plates and in gas combustion burners and also in thermal insulation and fire protection materials [2].

The conventional TE problem consists of a parabolic-type heat conduction equation and a hyperbolic-type wave equation. However, based on Fourier law, a portion of the solution of traditional heat conduction equation could extend to infinity. That is, once a homogeneous, isotropic elastic continuum is subjected to a thermal or mechanical disturbance, the effects of disturbance on temperature and displacement fields will be felt immediately at distances far from the source, which is against with practical physical observations [3]. Consequently, many generalized TE theories have been proposed based on the ''second sound'' effects in order to overcome such paradox, among which the Lord and Shulman (LS) [4] Green and Lindsay (GL) [5], and Green and Nagdhi (GN) theories [6], [7] are the most significant ones. The basic assumption of these theories is that the heat conduction phenomenon should be regarded as wave motion instead of diffusion. The GN theory without energy dissipation (GN-II theory) possesses several significant properties that differ from the development in TE material behavior of traditional classical theory. They are [7]: (a) it does not sustain energy dissipation; (b) the entropy flux vector in the theory is determined in terms of the same potential that also determines the stress; (c) it permits transmission of heat flow as thermal waves at finite speed.

TE Rayleigh wave is one of the most important topics in practical engineering fields due to the significance of thermoelasticity [8]. Chandrasekharaiah [9] utilized the generalized TE theory to analyze the phase velocity of Rayleigh wave by introducing two potential functions. However, the effects of thermal field were ignored in his report. Effects of rotation [10], viscoelasticity [11], piezoelectric [12], thermoelasticity [13], and magneto-thermo-elasticity [14] on the propagation of Rayleigh wave in half space have been presented by various researchers.

Recently, acoustic metamaterials and phononic crystals, which are usually comprised of two, or more various materials arranged periodically in space [15], have drawn much attention all over the world [16]-[18]. Band structure—the most notable properties of phononic crystals—can be utilized to describe the wave propagation in elastic mediums. A lot of theoretical and experimental papers [19]-[23] have been published to investigate such property. Some effective methods have been applied to calculate the band structure of phononic crystals, among which transfer matrix method (TFM), finite domain time difference (FDTD), plane wave expansion method (PWE), multi-scattering method (MST), and finite element method (FEM) are the most frequently-used. Acoustic metamaterial and phononic crystals have potential applications in vibration isolation [24], acoustic cloaking [25], and acoustic hyperlens [26] and so forth. Since the band gap phenomenon can be used to block certain frequencies, phononic crystals and acoustic metamaterials also should be a good choice for noise and vibration attenuation in thermal environments. Therefore, it is essential to investigate the TE coupling effects on the band structure of phononic crystals.

Investigations on the propagation of elastic surface waves in phononic crystals can be found in Ref [27]-[30]. Wu et al. [27] formulated the expression of bulk wave and surface wave via PWE method and investigated the band gap properties. The authors concluded that the mode can exchange suddenly around the sharp bend area. In a recent article [30], the concept of the localization factor was proposed to investigate the elastic Rayleigh elastic wave propagation and localization in disordered piezoelectric phononic crystals.

However, to the best of our knowledge, the band structure of TE Rayleigh wave in phononic crystals has not yet been studied in the literature. Different from previous studies of elastic surface wave propagation in phononic crystals, once phononic crystals are used in thermal environments, the thermoelastic coupling effects on band structure properties cannot be simply ignored. Since the wave and heat conduction equations are coupled, it is particularly worthy of academic investigation as well as practical application for the TE surface wave in phononic crystals. Based on GN-II theory, we proposed a generalized PWE method to investigate the TE buck wave band in 2-D and 3-D phononic crystals [31]. In this paper, we mainly focus on the band structure of TE Rayleigh surface waves in one-dimensional Aluminum/Epoxy phononic crystal under steady-state thermal environments. The expression of phase velocity and the generalized transfer matrix between two adjacent units is obtained by considering mechanical and thermal boundary conditions. Additionally, band structures corresponding to the real and imaginary wave vectors are calculated. The absolute value of the imaginary wave vector, which can be utilized to characterize the transmission property of the band gaps, is analyzed and compared with the case in pure elastic (PE) theory. The results presented in this paper are all compatible with and cross-checked by the existing results in the literature. Some new phenomena, as well as the possible reasons, are also discussed and analyzed.

## 2. Theory background

For a homogeneous and isotropic elastic solid, the basic equations for GN-II theory in the absence of body forces and heat resources are governed by [12]:

(i) Wave equation:
$$\sigma_{ij,j} = \rho \ddot{u}_i \quad (1)$$

(ii) Thermoelastic constitutive equation:
$$\sigma_{ij} = \lambda e_{kk}\delta_{ij} + 2\mu e_{ij} - mT\delta_{ij} \quad (2)$$

(iii) GN-II thermal conduction equation:
$$\kappa T_{,ii} = \rho c \ddot{T} + mT_0 \ddot{e} \quad (3)$$

where $i, j, k = 1, 2, 3$ refer to general coordinates. $\sigma_{ij}$ -the stress tensor, $\rho$ -mass density, $u_i$ -the displacement tensor, $e_{ij}$ -the strain tensor, $\delta_{ij}$ -the Kronecker tensor, $\lambda$ and $\mu$ - Lame constants of materials, $e$ -the dilatation, $T$ -temperature difference, $T_0$ -balanced temperature, $c$ -specific heat capacity, $\kappa$ -the material constant characteristic of the theory (not the coefficients of thermal conductivity, $\kappa = (\lambda + 2\mu)c/4$ implies the pure thermal wave happens to be slower than the longitudinal TE wave. [3]), $m = \alpha(3\lambda + 2\mu)$, $\alpha$ -the coefficient of thermal expansion. The invariant form of the wave equation can be expressed as

$$(\lambda+\mu)\nabla(\nabla\cdot \mathbf{u}) + \mu\nabla^2\mathbf{u} - mT_0\nabla\Theta = \rho\ddot{\mathbf{u}}$$
$$\rho\ddot{\Theta} + \bar{m}(\nabla\cdot\ddot{\mathbf{u}}) = \bar{\kappa}\nabla^2\Theta \quad (4)$$

where $\Theta = T/T_0$, $\bar{\kappa} = \kappa/c$, $\bar{m} = m/c$.

The 1-D phononic crystal composed of materials A and B are showed in Fig. 1. The thickness of materials A and B is $a_1$ and $a_2$, respectively. Then, the lattice constant is $a = a_1 + a_2$. The local coordinate $(x_j, y_j)$ of substructure is also illustrated in Fig. 1. For 1-D case, the components of displacement are $\mathbf{u} = (u_{jx}, u_{jy})$. The wave equation can be concisely expressed using tensor notation,

$$\rho_j u_{j\beta,tt} = (\lambda_j + \mu_j)u_{j\gamma,\gamma\beta} + \mu_j u_{j\beta,\gamma\gamma} - m_j T_0 \Theta_{j,\beta} \quad (5)$$

The generalized thermal conduction equation can be expressed as
$$\rho_j \Theta_{j,tt} + \bar{m}_j u_{j\beta,\beta tt} = \bar{\kappa}_j \Theta_{j,\beta\beta} \quad (6)$$

where $j = A, B$ denotes different layers, $\beta, \gamma = 1, 2$.

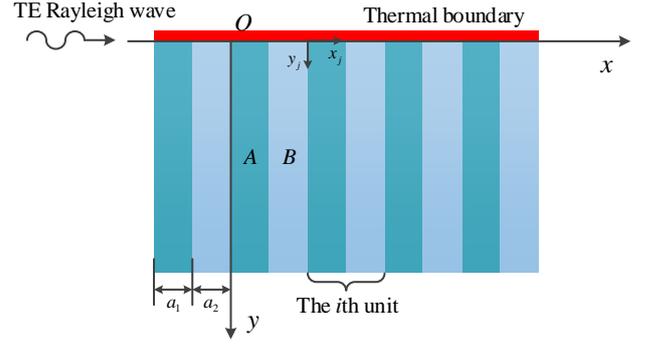

**Fig. 1.** Schematic diagram of TE Rayleigh wave in 1-D phononic crystal.

Solutions of the Eqs. (5) and (6) can be decomposed in terms of normal modes as follows,

$$\begin{aligned}
u_{jx}(x_j,y_j,t) &= A_j\exp(ik_j b_j y_j)\exp[i(k_j x_j - \omega t)] \\
&+ B_j\exp(ik_j b_j y_j)\exp[i(-k_j x_j - \omega t)] \\
u_{jy}(x_j,y_j,t) &= G_j\exp(ik_j b_j y_j)\exp[i(k_j x_j - \omega t)] \\
&+ H_j\exp(ik_j b_j y_j)\exp[i(-k_j x_j - \omega t)] \\
\Theta_j(x_j,y_j,t) &= P_j\exp(ik_j b_j y_j)\exp[i(k_j x_j - \omega t)] \\
&+ Q_j\exp(ik_j b_j y_j)\exp[i(-k_j x_j - \omega t)]
\end{aligned} \quad (7)$$

where $i = \sqrt{-1}$, $b_j$ denotes the decaying rate, $k_j$ represents the wave number in the $x$ direction. $A_j$, $B_j$, $G_j$, $H_j$, $P_j$, and $Q_j$ are the unknown magnitudes of the displacements and dimensionless temperature fields.

The decaying rate $b_j$ can be obtained only by considering the wave solution along the positive $x$-axis. The corresponding values $b_j$ determined by the solutions along the negative $x$-axis are the same as those of waves propagating along the positive $x$-axis. Substitute Eq. (7) into Eqs. (5) and (6) can yields

$$0 = \omega^2 \rho_j A_j - k_j^2(\lambda_j+2\mu_j)A_j - k_j^2 b_j^2 \mu_j A_j - k_j^2 b_j(\lambda_j+\mu_j)G_j$$
$$\quad -ik_j m_j T_0 P_j$$
$$0 = \omega^2 \rho_j G_j - k_j^2 b_j^2(\lambda_j+2\mu_j)G_j - k_j^2 \mu_j G_j - k_j^2 b_j(\lambda_j+\mu_j)A_j \quad (8)$$
$$\quad -ik_j b_j m_j T_0 P_j$$
$$0 = \omega^2 \rho_j P_j + ik_j \omega^2 \bar{m}_j A_j + ik_j b_j \omega^2 \bar{m}_j G_j - \bar{\kappa}_j k_j^2(1+b_j^2)P_j$$

Arrange Eq. (8) in the matrix form
$$\mathbf{RU} = \mathbf{0} \quad (9)$$

where $\mathbf{U} = \{A_j \; G_j \; P_j\}^T$, the elements of $\mathbf{R}$ are shown in Appendix A.

For the nontrivial solutions of $\mathbf{U}$, the determinants of the coefficient matrix $\mathbf{R}$ must vanish. It implies $|\mathbf{R}|=0$, which leads to

$$-b^6 c_{j\kappa}^2 c_{jt}^2 c_{jl}^2$$
$$+ b^4\left[\eta_j c_{j\Theta}^2 c_{jt}^2 c_{j\phi}^2 - 3c_{j\kappa}^2 c_{jt}^2 c_{jl}^2 + c_{j\phi}^2 c_{j\kappa}^2(c_{jt}^2+c_{jl}^2) + c_{j\phi}^2 c_{jt}^2 c_{jl}^2\right]$$
$$+ b^2\begin{bmatrix} 2\eta_j c_{j\phi}^2 c_{j\Theta}^2 c_{jt}^2 - \eta_j c_{j\Theta}^2 c_{j\phi}^4 - 3c_{j\kappa}^2 c_{jt}^2 c_{jl}^2 \\ +2c_{j\phi}^2 c_{j\kappa}^2(c_{jt}^2+c_{jl}^2) + 2c_{j\phi}^2 c_{jt}^2 c_{jl}^2 - c_{j\phi}^4(c_{j\kappa}^2+c_{jt}^2+c_{jl}^2) \end{bmatrix} \quad (10)$$
$$+ (c_{jt}^2 - c_{j\phi}^2)\left[\eta_j c_{j\phi}^2 c_{j\Theta}^2 - c_{j\kappa}^2 c_{jl}^2 + c_{j\phi}^2(c_{j\kappa}^2+c_{jl}^2) - c_{j\phi}^4\right] = 0$$

where $c_{jt} = \sqrt{\mu_j/\rho_j}$ —the velocity of transversal wave, $c_{jl} = \sqrt{(\lambda_j+2\mu_j)/\rho_j}$ —the velocity of longitudinal wave,

$c_{j\Theta} = \sqrt{m_j T_0 / \rho_j}$ —the velocity related to volume dilatation due to the thermal expansion, $c_{j\phi} = \omega/k_j$ —the phase velocity of the $j$-th substructure, $c_{j\kappa} = \sqrt{\bar{\kappa}_j / \rho_j}$ —the velocity of generalized thermal wave, $\eta_j = \bar{m}_j / \rho_j$ —the dimensionless variable.

If the TE coupling are not included, Eq. (10) can be reduced to the following well known expression

$$\left[c_{jl}^2 b^2 - (c_{jl}^2 - c_{j\phi}^2)\right]\left[c_{jt}^2 b^2 - (c_{jt}^2 - c_{j\phi}^2)\right] = 0 \quad (11)$$

The equation is exactly the same as what we can easily find in classical Rayleigh wave theory which indicates Eq. (10) is compatible with classical PE theory. There exist three roots for $\text{Im}(b) > 0$ of equation (10), but the condition must be satisfied, which indicates that both the displacements and the temperature difference diminish along the positive y-axis. According to Eq. (9), the relations of the unknown magnitudes can be expressed as

$$\begin{aligned}G_{jn} &= f_{jn} A_{jn}, \quad H_{jn} = -f_{jn} B_{jn}, \quad (n=1,2,3) \\ P_{jn} &= g_{jn} A_{jn}, \quad Q_{jn} = -g_{jn} B_{jn}, \quad (n=1,2,3)\end{aligned} \quad (12)$$

The corresponding coefficients are showed in the Appendix. $n$ denotes three qualified roots of $b$. Thus, the general solutions for the displacements and temperature difference are written as

$$\begin{aligned}u_{jx}(x_j, y_j, t) &= \left[\sum_{n=1}^{3} A_{jn} \exp(i k_j b_{jn} y_j)\right] \exp[i(k_j x_j - \omega t)] \\ &+ \left[\sum_{n=1}^{3} B_{jn} \exp(i k_j b_{jn} y_j)\right] \exp[i(-k_j x_j - \omega t)] \\ u_{jy}(x_j, y_j, t) &= \left[\sum_{n=1}^{3} f_{jn} A_{jn} \exp(i k_j b_{jn} y_j)\right] \exp[i(k_j x_j - \omega t)] \\ &- \left[\sum_{n=1}^{3} f_{jn} B_{jn} \exp(i k_j b_{jn} y_j)\right] \exp[i(-k_j x_j - \omega t)] \\ \Theta_j(x_j, y_j, t) &= \left[\sum_{n=1}^{3} g_{jn} A_{jn} \exp(i k_j b_{jn} y_j)\right] \exp[i(k_j x_j - \omega t)] \\ &- \left[\sum_{n=1}^{3} g_{jn} B_{jn} \exp(i k_j b_{jn} y_j)\right] \exp[i(-k_j x_j - \omega t)]\end{aligned} \quad (13)$$

Consider the following mechanical and thermal boundary conditions: (i) Mechanical boundary condition: $y = 0$, the stresses are assumed to be free: $\sigma_{jy}|_{y=0} = 0$, $\tau_{jyx}|_{y=0} = 0$. (ii) Thermal boundary: $y = 0$: $\Theta_{j,y} + h_j \Theta_j = 0$. where $h_j$ is a constant. If the boundary is thermally insulated, $h_j = 0$. And if the boundary is maintained in an isothermal state, $h_j \to \infty$. Substitute Eq. (13) into the boundary conditions, a set of linear equations of $A_{jn}$ are given by

$$\begin{aligned}0 &= \sum_{n=1}^{3}\left[i(\lambda_j + 2\mu_j) f_{jn} k_j b_{jn} + i k_j \lambda_j - m_j T_0 g_{jn}\right] A_{jn} \\ 0 &= \sum_{n=1}^{3}(i \mu_j k_j b_{jn} + i \mu_j f_{jn} k_j) A_{jn} \\ 0 &= \sum_{n=1}^{3} g_{jn}(i k_j b_{jn} + h_j) A_{jn}\end{aligned} \quad (14)$$

To obtain the nontrivial solutions of the amplitudes $A_{jn}$, the determinants of the coefficient matrix equals zero which implies that

$$|\mathbf{S}| = 0 \quad (15)$$

where the elements of $\mathbf{S}$ are listed in Appendix A.

Consequently, the phase speed equation for Rayleigh waves in the context of GN-II theory can be eventually expressed as

$$\begin{aligned}&(2 - c_{j\phi}^2 / c_{jt}^2)^2 = \\ &\frac{4\sqrt{1 - c_{j\phi}^2 / c_{jt}^2}\left[\sqrt{(1 - c_{j\phi}^2 / V_{j1}^2)(1 - c_{j\phi}^2 / V_{j2}^2)} - N_{j1} h_j / k_j\right]}{N_{j2} - h_j / k_j}\end{aligned} \quad (16)$$

We can also obtain the relationship between $A_{jn}$, that is $A_{j2} = q_{j2} A_{j1}$, $A_{j3} = q_{j3} A_{j1}$. $q_{j2}$, $q_{j3}$, and some intermediate components appeared in Eq. (16) are all listed in Appendix A. The similar phase speed equation can also be derived by introducing two displacement potential functions [9]. Once the phase speed is determined, the value of $b^2$ can be derived accordingly. As $\varepsilon_{jT}$ couples the equations corresponding to the elastic wave propagation and heat conduction, it can be called the coupling parameter. $\varepsilon_{jT} = 0$ corresponds to the uncoupled case. The dimensionless coupling parameter is influenced by the material constants and the balanced temperature. Since the magnitude of the thermal expansion is approximately at $10^{-6}$, a strong coupling parameter is hard to realize. The coupling parameter for traditional materials ranges from 0.01 to 0.1[9]. It is worthy to point out that once the couplings between the thermal and displacement fields are ignored ($\varepsilon_T \to 0$), the phase speed Eq. (16) will reduce to

$$(2 - c_{j\phi}^2 / c_{jt}^2)^2 = 4\sqrt{(1 - c_{j\phi}^2 / c_{jl}^2)(1 - c_{j\phi}^2 / c_{jt}^2)} \quad (17)$$

It can be immediately noted that this is the phase speed equation for Rayleigh waves in the classical theory of elasticity. A significantly common feature of Eq. (16) (with $h = 0$ or $h \to \infty$) and (19) is that these two equations are independent of the wave vector $k_j$, which predicts a non-dispersive property for Rayleigh waves. This is not the case in the conventional thermoelasticity theory, or the LS and GL theories, in which the Rayleigh waves do experience dispersion behavior [33], [34]. The difference of Rayleigh waves between the GN theory and the other three thermoelasticity theories mainly stems from the fact that the GN theory does not accommodate energy dissipation [9].

According to the boundary conditions at the left and right sides of the two sub-cells in the $i$-th unit cell and based on the periodicity of material constants and Bloch–Floquet theorem, the standard eigenvalue equation can be formulated as

$$|\mathbf{T}_i - e^{ika} \mathbf{I}_6| = 0 \quad (18)$$

in which $\mathbf{T}_i$ is the transfer matrix (see appendix), $\mathbf{I}_6$ is the $6 \times 6$ identity matrix. By solving Eq. (18) for $\omega$ as a function of $k$ in the first Brillouin zone, the band structures of generalized TE Rayleigh wave in phononic crystals can be built eventually.

In addition, for any given $\omega$, if the reduced wave vector $k$ belongs to the solutions of Eq. (18), therefore $-k$ should also be one of the solutions which corresponds to the wave moving to the negative $x$-axis. Actually, it can be proved that $e^{ika}$ always accompany with $e^{-ika}$ for the transfer matrix, which means the its eigenvalues should satisfy the following condition [35]

$$\ln(\lambda_1) \geq \ln(\lambda_2) \geq \ln(\lambda_3) \geq 0 \geq \ln(\lambda_4)$$
$$(=-\ln(\lambda_3)) \geq \ln(\lambda_5)(=-\ln(\lambda_2)) \geq \ln(\lambda_6)(=-\ln(\lambda_1)) \quad (19)$$

It can be observed from Eq. (19) that two pairs of eigenvalues denote the PE parts, the rest are the additional thermal wave part. Set $k = k_{re} + ik_{im}$ and $k = -(k_{re} + ik_{im})$, we can immediately obtain $|k_{im}| = |\ln(\lambda_i)|/a$. For any given $\omega$, if $k_{im} = 0$, the frequency lies in the pass band region. Otherwise, if $k_{im} \neq 0$, $\omega$ is in the forbidden band field. The absolute value of the imaginary wave vector can be employed to describe the attenuation character of the band gap [36]. The larger the value is, the stronger the attenuation will be.

## 3. Results and discussion

The material constants are listed in table 1 as well as the corresponding coupling parameters. From the table, we can figure out that the coupling parameters of the two materials are small enough that indicates a weak coupling between the elastic and thermal equations. The lattice constant is set to be $a = 0.15$ m. The filling fraction ratio $f = a_1/a$ is taken as 0.5. The normalized frequency is $\omega a/2\pi c_t$ in which $c_t$ represents the transversal velocity of the Epoxy layer. The balanced temperature $T_0$ is set to be 353K (80℃). The constant $h = 0$ implies a thermal insulating boundary condition.

**Table 1.** Material constants.

| Materials | | Aluminium | Epoxy |
| --- | --- | --- | --- |
| Lame constant $\lambda$ (Pa) | | 6.82E10 | 4.43E9 |
| Shear mode $\mu$ (Pa) | | 2.87E10 | 1.59E9 |
| Mass density $\rho$ (kg/m³) | | 2.73E3 | 1.18E3 |
| Heat coefficient $k_T$ (W/(m·K)) | | 230 | 0.2 |
| Specific capacity $c$ (J/(kg·K)) | | 900 | 550 |
| Dilatation coefficient $\alpha$ (1/K) | | 23.2E-6 | 67.1E-6 |
| Coupling parameter $\varepsilon_T$ | | 0.0096 | 0.0198 |
| Thermal wave velocity (m/s) | | 3391.5 | 1269.7 |
| Decaying rates $b$ | TE | 0.3550i, 0.8960i, 0.4375i | 0.3498i, 0.9063i, 0.4973i, |
| | PE | 0.3539i; 0.8945i | 0.3476i, 0.9035i |
| Phase velocity (m/s) | TE | 3030.5 | 1087.4 |
| | PE | 3032.6 | 1088.4 |

The decaying rates of Aluminum and Epoxy are also illustrated in table 1. It can be observed from the table that all the decaying rates satisfy the requirement $\text{Im}(b) > 0$. And the first two decaying rates considering the TE coupling effects are almost consistent with those in PE theory. The third value emerges mainly because of the introduced thermoelastic coupling effects. In addition to the decaying rates, the phase velocities of Rayleigh wave in TE theory are both smaller than those in PE theory.

In Fig. 2, the blue lines and red hollow dots denote TE wave bands and PE wave bands, respectively. We can find that besides the PE bands that appear in the TE wave band structure, three additional thermal wave bands also appear in the high and low-frequency regions, which do not exist in the PE medium. Yu et al. [37] pointed out that the phase velocities of thermal modes decrease with the increase of frequency, and eventually congregate together with the elastic modes. Thus, the existence of the thermal waves is much dependent on the hyperbolic-type heat conduction equation.

Once the TE couplings are included, the thermal bands can share similar behaviors as the elastic bands. The band structure of TE Rayleigh wave is comprised of thermal wave and elastic wave bands.

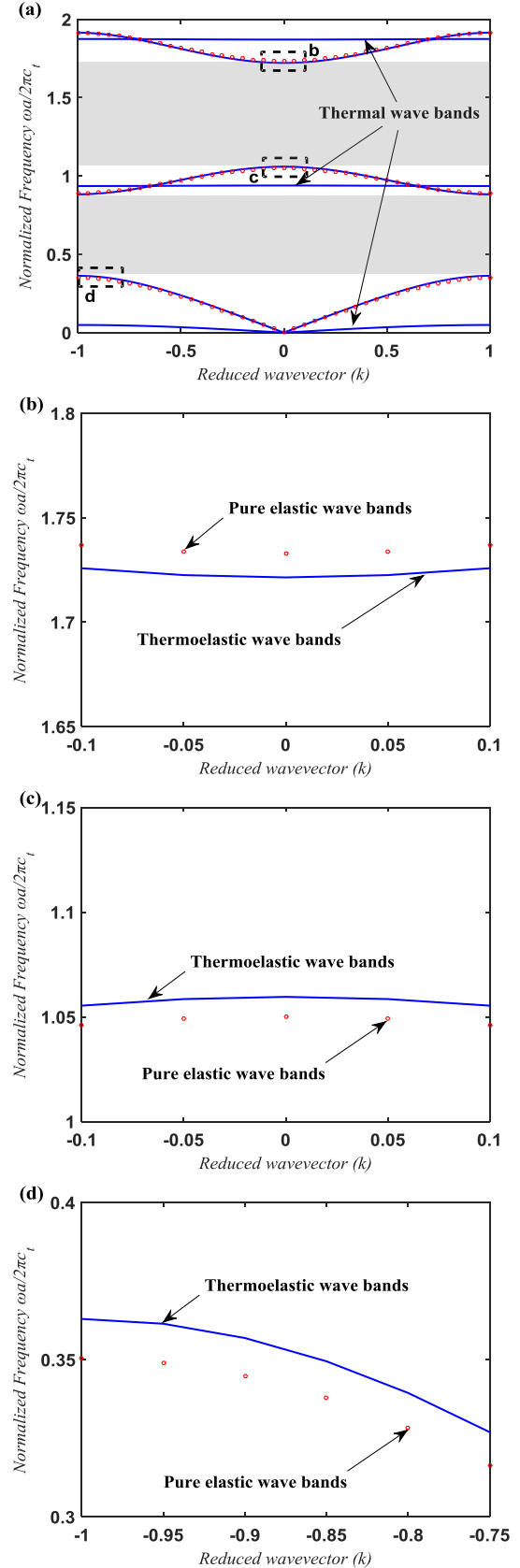

**Fig. 2.** Band structure of generalized TE Rayleigh wave in phononic crystals. The filling fraction ratio is 0.5. (a) The band structure. (b)(c)(d) Enlarged drawing corresponding to the b, c, d sections in (a).

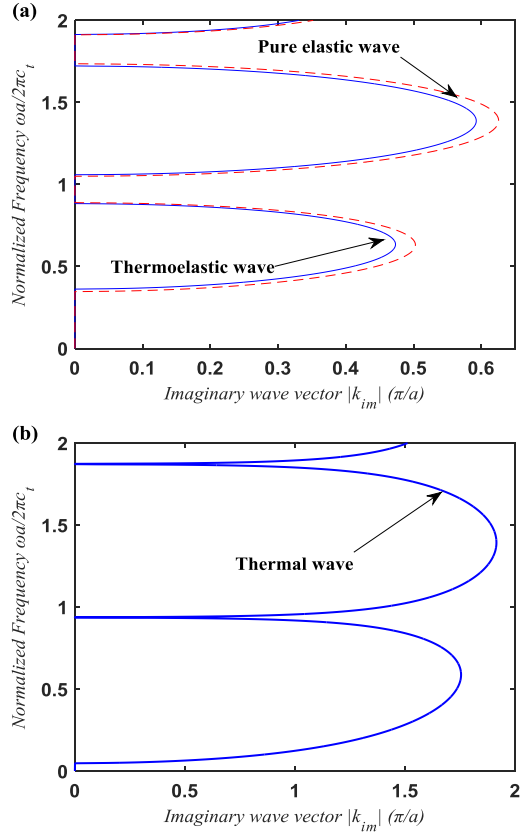

**Fig. 3.** The band structure corresponds to the imaginary wave vector. (a) Band structure of elastic wave in TE and PE theory. (b) The additional thermal wave bands of TE Rayleigh wave.

There are two complete gaps in Fig. 2 which shows that any Rayleigh waves with the frequency lie in the region cannot propagate across the structure. However, it is worth to point out that there exist some differences between the band structures of TE Rayleigh wave and PE Rayleigh wave, which demonstrates that the thermoelasticity does influence the originally PE wave bands. One interesting difference is that the upper and lower boundaries of each band gap of the pure thermal wave are modified once the coupling effects are taken into account. From the enlarged figures, we can further observe that the TE coupling can narrow down the band gap widths of the pure thermal wave bands. Besides the effect of TE coupling on boundaries of band gaps, it can also affect the transmission ability of the band gap. In Fig. 3(a), the imaginary wave vector is employed to characterize the transmission properties of the band gaps. It can be seen that the transmission abilities of TE Rayleigh's first two band gaps are both smaller than those of PE Rayleigh band gaps. In addition, in Fig. 3(b) we are also informed that the imaginary wave vector of thermal wave is larger than the elastic wave. All the intersection points on the vertical axis in Fig. 2 are consistent with the corresponding points in Fig. 2. It is worth mentioning that except at low frequency, there are nearly no pass bands of the pure thermal wave in phononic crystals which demonstrates the thermal wave is more likely to disappear when compared with elastic waves.

The displacements of each layer versus the depth at 500Hz, 1000Hz, 1500Hz and 2000Hz are depicted in Fig. 4. It can be observed that the magnitudes of either displacement decline exponentially as depth increases. Additionally, the displacement in Aluminum layer decrease a little slowly than in Epoxy layer, which is due to the differences between decaying rates in each layer. We are also informed that as frequency increases, the waves are more likely to be located on the surfaces instead of propagating in the depth direction.

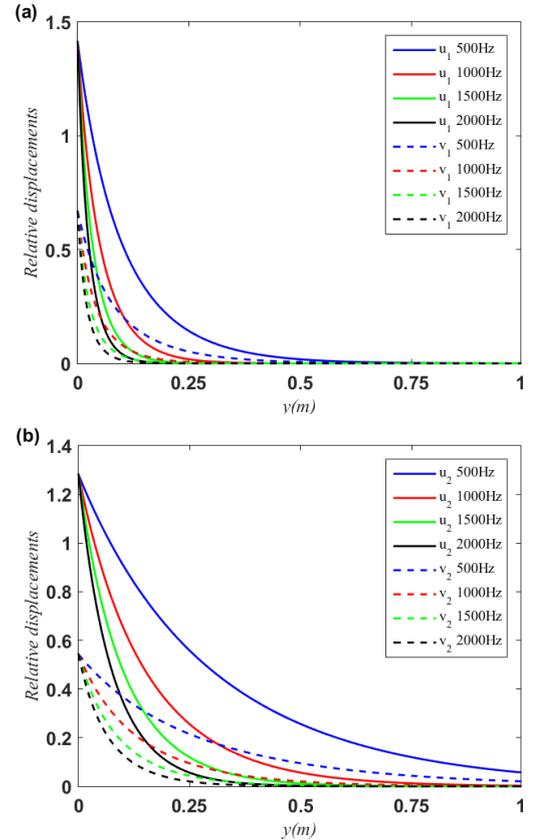

**Fig.4.** The relative displacements of each layer versus the depth at various frequencies. (a) Displacements in Epoxy layers. (b) Displacements in Aluminum layer.

After investigating the numerical value of the coupling elements in Eq. (6), we find out that the orders of magnitude of $\bar{m}_j$ is about $10^4$ smaller than the other elastic material constants, which leads to the influence of thermoelasticity is limited. In addition, the non-dimensional temperature differences of Aluminum and Epoxy layers are small enough (see Fig. 5) compared with the balanced temperature. It can be viewed from the figure that the temperature difference is located on the surface and is decaying along the positive $y$-axis. And the temperature difference is closely related to the normalized frequency. When frequency increases, the temperature enlarges correspondingly, but its localization on the surface becomes more serious. The low magnitude of the temperature difference implies that the produced thermal stresses is relatively small in the whole structure. However, from the above observation, we can figure out that thermal stresses do influence the band structure of phononic crystals even it is relatively small. Therefore, effects of TE coupling on the band structure of phononic crystals are much dependent on thermal stresses. If we apply sensible coupling effects, typical of some non-traditional materials, such as certain high-performance composites, and preset a large temperature gradient in the structure, the effect of thermal stresses on the band structure may be much more remarkable. Another possible reason is that the ideally periodic boundary conditions are suitable for infinite structures. Once a phononic-crystal plate with finite size is embedded in thermal environments, the effect of thermal stresses will occur if the expansion or compression cannot happen freely because of boundary and internal constraints. The effects of thermal stresses cannot be ignored since the magnitude of thermal stresses is sometimes comparable to the elastic constants.

It is also noticeable to point out that the band gap cannot confine heat energy or forbid heat transfer since the thermal field is assumed to be steady and the element $\Theta_j^{(i)}$ in $\mathbf{V}_j^{(i)}$ only represents the dimensionless temperature difference. While the absolute temperature is comprised of temperature difference and the reference temperature.

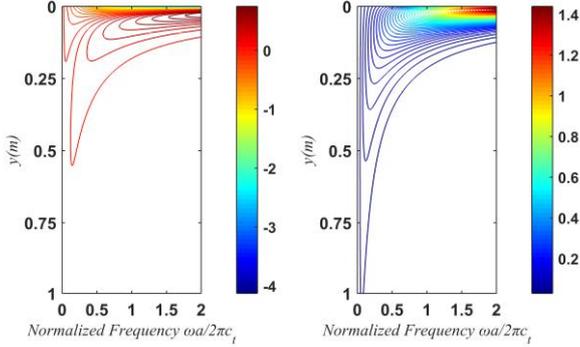

**Fig. 5.** The dimensionless temperature differences $\Theta$ of each layer versus the depth and normalized frequency. Left: Epoxy. Right: Aluminium.

The filling fraction has long been known to influence the band gaps of phononic crystals. In this section, we will follow the same procedure as the literature to explore the effect of TE coupling on band gap at various filling ratios.

The lower and upper boundaries of the first and second band gap versus the filling ratio $f$ are showed in Fig. 6. It can be observed that the PE and TE band gap boundaries possess the same trend. For the first TE band gap LBF, it initially declines slightly from 0.4085 at $f$=0.1 to 0.3595 at $f$=0.4, then the value goes up to 0.5994 when $f$=0.9. It is noticeable to point out that the counterpart of the first PE band gap LBF is a little lower throughout the whole range of filling ratio. The value of the first TE band gap UBF climbs rapidly until it reaches a peak at 1.412 when $f$=0.8. After that point, it starts to fall off. Unlike the LBF of the first PE band gap, the UBF is appreciably higher than that of the first TE band gap UBF. The LBF of the second TE band gap, first drops then increases promptly to a peak value 1.705 at $f$=0.8. Then the value declines from $f$=0.8 to 0.9. The lower boundary frequencies of the second PE wave band are also lower than the second TE band gap LBF. The general tendency for the second TE band gap UBF is rising except a fluctuation between $f$=0.6 and 0.7. The corresponding values of PE band gap UBF are higher than its counterparts. The width of the first TE band gap reaches the maximum point 0.9472 when $f$=0.8. But for width of the second TE band gap, there exists a minimum value when $f$=0.7.

From the above observations and discussions, we are informed by the fact that the PE band gap boundaries are influenced by the TE coupling effects. This phenomenon can be further understood by the curve of the normalized band gap width in Fig. 6. It is a noteworthy phenomenon that the widths of PE band gaps are larger than that of the TE band gaps, which demonstrates that the TE coupling has the ability to narrow down the widths of PE band gaps especially when the filling ratio is around 0.8.

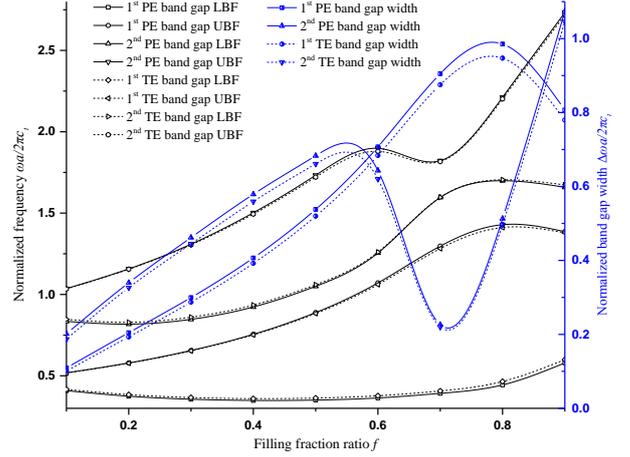

**Fig. 6.** Lower and upper boundaries of the first and second band gaps and the corresponding normalized widths of band gap versus filling ratio $f$. LBF and UBF represent the lower and upper boundary frequencies of the band gap.

## 4. Conclusion

In summary, we have investigated the band structure of TE Rayleigh wave in one-dimensional phononic crystal in the context of GN-II theory by taking into account the TE coupling effects. From the theoretical and numerical results, the following conclusions can be drawn: (1) A new finding reported in this paper is that the generalized band structure of TE Rayleigh waves is comprised of the elastic wave bands and thermal wave bands. Thermal bands have similar behaviors as the elastic bands, but the lifetime of thermal modes is very short since the attenuation rates of the thermal wave mode are larger than the elastic mode. (2) Thermoelasticity influences the band-gap width of TE Rayleigh wave in phononic crystals. Based on Aluminum/Epoxy 1-D phononic crystals in this report, TE coupling effects can narrow down the band gap width especially when the filling ratio is around 0.8. (3) The attenuation property of the corresponding band gap is also affected by the TE coupling behavior besides the width of the band gap. For Aluminum/Epoxy 1-D phononic crystals, once the coupling is included, the attenuation of the band gaps will be smaller than the counterpart of the PE band gaps. These novel phenomena and findings do provide an approach to investigate how TE coupling effects can influence the band structure, which may have important implications for acoustic attenuation.

## Acknowledgements

This work is supported by the Harbin Institute of Technology and the China Scholarship Council (Grant No. 201606120065).

## Appendix

The elements of matrix $\mathbf{R}$ in Eq. (9)

$$\begin{aligned}
&R_{11} = \omega^2 \rho_j - k_j^2(\lambda_j + 2\mu_j) - k_j^2 b_j^2 \mu_j; \quad R_{12} = -k_j^2 b_j(\lambda_j + \mu_j) \\
&R_{13} = -ik_j m_j T_0; \quad R_{21} = -k_j^2 b_j(\lambda_j + \mu_j) \\
&R_{22} = \omega^2 \rho_j - k_j^2 b_j^2(\lambda_j + 2\mu_j) - k_j^2 \mu_j \\
&R_{23} = -ik_j b_j m_j T_0; \quad R_{31} = ik_j \omega^2 \bar{m}_j; \quad R_{32} = ik_j b_j \omega^2 \bar{m}_j \\
&R_{33} = \omega^2 \rho_j - \bar{\kappa}_j k_j^2(1 + b_j^2)
\end{aligned} \quad (A.1)$$

The coefficients of Eq. (12)

$$f_{j1} = b_{j1}$$
$$g_{j1} = \mathrm{i}k_j\left[(b_{j1}^2+1)c_{jl}^2 - c_{j\phi}^2\right]/c_{j\Theta}^2$$
$$f_{j2} = \frac{-(c_{jl}^2 - c_{jt}^2)b_{j2}\left[(b_{j2}^2+1)c_{j\kappa}^2 - c_{j\phi}^2\right] + \eta_j c_{j\Theta}^2 c_{j\phi}^2 b_{j2}}{\left[c_{j\kappa}^2 b_{j2}^2 - (c_{j\phi}^2 - c_{j\kappa}^2)\right]\left[c_{jl}^2 b_{j2}^2 - (c_{j\phi}^2 - c_{jt}^2)\right] - \eta_j c_{j\Theta}^2 c_{j\phi}^2 b_{j2}^2}$$
$$g_{j2} = \frac{\mathrm{i}\eta_j k_j c_{j\phi}^2\left[(b_{j2}^2+1)c_{jt}^2 - c_{j\phi}^2\right]}{\left[c_{j\kappa}^2 b_{j2}^2 - (c_{j\phi}^2 - c_{j\kappa}^2)\right]\left[c_{jl}^2 b_{j2}^2 - (c_{j\phi}^2 - c_{jt}^2)\right] - \eta_j c_{j\Theta}^2 c_{j\phi}^2 b_{j2}^2}$$
$$f_{j3} = \frac{\left[c_{j\kappa}^2 b_{j3}^2 - (c_{j\phi}^2 - c_{j\kappa}^2)\right]\left[c_{jl}^2 b_{j3}^2 - (c_{j\phi}^2 - c_{jl}^2)\right] - \eta_j c_{j\Theta}^2 c_{j\phi}^2}{-(c_{jl}^2 - c_{jt}^2)b_{j3}\left[(b_{j3}^2+1)c_{j\kappa}^2 - c_{j\phi}^2\right] + \eta_j b_{j3} c_{j\Theta}^2 c_{j\phi}^2}$$
$$g_{j3} = \frac{\mathrm{i}\eta_j k_j c_{j\phi}^2\left[(b_{j3}^2+1)c_{jt}^2 - c_{j\phi}^2\right]}{-(c_{jl}^2 - c_{jt}^2)b_{j3}\left[(b_{j3}^2+1)c_{j\kappa}^2 - c_{j\phi}^2\right] + \eta_j c_{j\Theta}^2 c_{j\phi}^2 b_{j3}}$$

(A.2)

The relationship between $A_{jn}$

$$q_{j2} = -\frac{\left\{(\lambda_j + 2\mu_j)\left[b_{j1}b_{j3}(f_{j1} - f_{j3}) + f_{j1}f_{j3}(b_{j1} - b_{j3})\right] - \lambda_j(b_{j1} - b_{j3} + f_{j1} - f_{j3})\right\}k_j - \mathrm{i}m_j T_0\left[g_{j3}(b_{j1} + f_{j1}) - g_{j1}(b_{j3} + f_{j3})\right]}{\left\{(\lambda_j + 2\mu_j)\left[b_{j2}b_{j3}(f_{j2} - f_{j3}) + f_{j2}f_{j3}(b_{j2} - b_{j3})\right] - \lambda_j(b_{j2} - b_{j3} + f_{j2} - f_{j3})\right\}k_j - \mathrm{i}m_j T_0\left[g_{j2}(b_{j3} + f_{j3}) - g_{j3}(b_{j2} + f_{j2})\right]}$$

$$q_{j3} = \frac{\left\{(\lambda_j + 2\mu_j)\left[b_{j1}b_{j2}(f_{j1} - f_{j2}) + f_{j1}f_{j2}(b_{j1} - b_{j2})\right] - \lambda_j(b_{j1} - b_{j2} + f_{j1} - f_{j2})\right\}k_j - \mathrm{i}m_j T_0\left[g_{j1}(b_{j2} + f_{j2}) - g_{j2}(b_{j1} + f_{j1})\right]}{\left\{(\lambda_j + 2\mu_j)\left[b_{j2}b_{j3}(f_{j2} - f_{j3}) + f_{j2}f_{j3}(b_{j2} - b_{j3})\right] - \lambda_j(b_{j2} - b_{j3} + f_{j2} - f_{j3})\right\}k_j - \mathrm{i}m_j T_0\left[g_{j2}(b_{j3} + f_{j3}) - g_{j3}(b_{j2} + f_{j2})\right]}$$

(A.3)'

The components of **S** in Eq. (15)

$$S_{11} = \mathrm{i}k_j(\lambda_j + 2\mu_j)f_{j1}b_{j1} + \mathrm{i}k_j\lambda_j - m_j T_0 g_{j1}$$
$$S_{12} = \mathrm{i}k_j(\lambda_j + 2\mu_j)f_{j2}b_{j2} + \mathrm{i}k_j\lambda_j - m_j T_0 g_{j2}$$
$$S_{13} = \mathrm{i}k_j(\lambda_j + 2\mu_j)f_{j3}b_{j3} + \mathrm{i}k_j\lambda_j - m_j T_0 g_{j3}$$
$$S_{21} = \mathrm{i}k_j(b_{j1} + f_{j1});\ S_{22} = \mathrm{i}k_j(b_{j2} + f_{j2})$$
$$S_{23} = \mathrm{i}k_j(b_{j3} + f_{j3});\ S_{31} = g_{j1}(\mathrm{i}k_j b_{j1} + h_j)$$
$$S_{32} = g_{j2}(\mathrm{i}k_j b_{j2} + h_j);\ S_{33} = g_{j3}(\mathrm{i}k_j b_{j3} + h_j)$$

(A.4)

The components of Eq. (16)

$$N_{j1} = M_{j1}\sqrt{1 - c_{j\phi}^2/V_{j1}^2} + M_{j2}\sqrt{1 - c_{j\phi}^2/V_{j2}^2}$$
$$N_{j2} = M_{j2}\sqrt{1 - c_{j\phi}^2/V_{j1}^2} + M_{j1}\sqrt{1 - c_{j\phi}^2/V_{j2}^2}$$
$$M_{j1} = V_{j1}^2(1 - V_{j2}^2/c_{jl}^2)/\Delta_j$$
$$M_{j2} = -V_{j2}^2(1 - V_{j1}^2/c_{jl}^2)/\Delta_j$$
$$V_{j1} = \sqrt{[c_{j\kappa}^2 + (1+\varepsilon_T)c_{jl}^2 + \Delta_j]/2}$$
$$V_{j2} = \sqrt{[c_{j\kappa}^2 + (1+\varepsilon_T)c_{jl}^2 - \Delta_j]/2}$$
$$\Delta_j = \sqrt{[c_{j\kappa}^2 - (1+\varepsilon_{jT})c_{jl}^2]^2 + 4\varepsilon_{jT} c_{j\kappa}^2 c_{jl}^2}$$
$$\varepsilon_{jT} = m_j^2 T_0/[\rho_j c_j(\lambda_j + 2\mu_j)]$$

(A.5)

Transfer matrix in Equation (18)

$$\mathbf{T}_i = \mathbf{T}_{Btr}^{(i)}\mathbf{T}_{Atr}^{(i)}$$

(A.6)

where: $\mathbf{T}_{jtr}^{(i)} = \mathbf{T}_{jL}^{(i)}\mathbf{\Lambda}_j[\mathbf{T}_{jL}^{(i)}]^{-1}$

$$\mathbf{\Lambda}_j = \mathrm{diag}\left[e^{\mathrm{i}\omega\alpha_j/c_{j\phi}}\ e^{\mathrm{i}\omega\alpha_j/c_{j\phi}}\ e^{\mathrm{i}\omega\alpha_j/c_{j\phi}}\ e^{-\mathrm{i}\omega\alpha_j/c_{j\phi}}\ e^{-\mathrm{i}\omega\alpha_j/c_{j\phi}}\ e^{-\mathrm{i}\omega\alpha_j/c_{j\phi}}\right]$$

$$T_{jL11} = T_{jL12} = T_{jL13} = T_{jL14} = T_{jL15} = T_{jL16} = 1$$
$$T_{jL21} = -T_{jL24} = f_{j1};\ T_{jL22} = -T_{jL25} = f_{j2}$$
$$T_{jL23} = -T_{jL26} = f_{j3};\ T_{jL31} = -T_{jL34} = g_{j1}$$
$$T_{jL32} = -T_{jL35} = g_{j2};\ T_{jL33} = -T_{jL36} = g_{j3}$$
$$T_{jL41} = -T_{jL44} = \mathrm{i}\omega(\lambda_j + 2\mu_j)/c_{j\phi} + \mathrm{i}\omega f_{j1}b_{j1}\lambda_j/c_{j\phi} - m_j T_0 g_{j1}$$
$$T_{jL42} = -T_{jL45} = \mathrm{i}\omega(\lambda_j + 2\mu_j)/c_{j\phi} + \mathrm{i}\omega f_{j2}b_{j2}\lambda_j/c_{j\phi} - m_j T_0 g_{j2}$$
$$T_{jL43} = -T_{jL46} = \mathrm{i}\omega(\lambda_j + 2\mu_j)/c_{j\phi} + \mathrm{i}\omega f_{j3}b_{j3}\lambda_j/c_{j\phi} - m_j T_0 g_{j3}$$
$$T_{jL51} = T_{jL54} = \mathrm{i}\omega\mu_j(f_{j1} + b_{j1})/c_{j\phi};\ T_{jL52} = T_{jL55} = \mathrm{i}\omega\mu_j(f_{j2} + b_{j2})/c_{j\phi}$$
$$T_{jL53} = T_{jL56} = \mathrm{i}\omega\mu_j(f_{j3} + b_{j3})/c_{j\phi}$$
$$T_{jL61} = T_{jL64} = \mathrm{i}\omega g_{j1}k_{Tj}T_0/c_{j\phi};\ T_{jL62} = T_{jL65} = \mathrm{i}\omega g_{j2}k_{Tj}T_0/c_{j\phi}$$
$$T_{jL63} = T_{jL66} = \mathrm{i}\omega g_{j3}k_{Tj}T_0/c_{j\phi}$$

**References**

[1] J L. Nowinski Theory of thermoelasticity with applications. Sijthoff & Noordhoff International Publishers, 1978.
[2] N. Sarkar and A. Lahiri 2012 International Journal of Engineering Science 51 310.
[3] H. Al-Qahtani, S.K. Datta 2004 J. Appl. Phys. **96** 3645.
[4] H.W. Lord, Y. Shulman 1967 J. Mech. Phys. Solids **15** 299.
[5] A. E. Green, K.A. Lindsay 1972 J. Elasticity **2** 1.
[6] A.E. Green, P.M. Naghdi 1992 J. Therm. Stresses **15** 253.
[7] A.E. Green, P.M. Naghdi 1993 J. Elasticity **31** 189.
[8] T.P. Ivanov, R. Savova 2010 Int. J. Solids Struct. **47** 1972.
[9] D.S. Chandrasekharaiah 1997 Mech. Res. Commun. **24** 93.
[10] J.N. Sharma, V. Walia 2007 Int. J. Solids Struct. **44** 1060.
[11] J.N. Sharma, M.I.A. Othman 2007 Int. J. Solids Struct. **44** 4243.
[12] J.N. Sharma, V. Walia, S. K Gupta 2008 Int. J. Mech. Sci. **50** 433.
[13] B. Singh, R. Pal 2011 Applied Mathematics **02** 521.
[14] A. Abd-Alla, S. Abo-Dahab, F. Bayones 2012 J. Vib. Control **19** 1395.
[15] M.I. Hussein, M.J. Leamy, M. Ruzzene 2014 Appl. Mech. Rev. **66** 040802.
[16] Z. Liu, X. Zhang, Y. Mao, Y. Zhu, Z. Yang, C.T. Chan, P. Sheng 2000 Science **289** 1734.
[17] Y.-Z. Wang, F.-M. Li, K. Kishimoto, Y.-S. Wang, W.-H. Huang Wave Motion **46** 47.
[18] D. Torrent and José Sánchez-Dehesa 2007 New J. Phys. **9** 323.
[19] D. García-Pablos, M. Sigalas, F. R. Montero de Espinosa, M. Torres, M. Kafesaki, N. García 2000 Phys. Rev. Lett. **84** 4349.
[20] T.-T. Wu, Z.-C. Hsu, Z.-G. 2005 Phys. Rev. B **71** 064303.
[21] T.-T. Wu, L.-C. Wu, Z.-G. Huang 2005 J. Appl. Phys. **97** 094916.
[22] J.F. Robillard, O.B. Matar, J.O. Vasseur, P.A. Deymier, M. Stippinger, A.C. Hladky-Hennion, Y. Pennec, B. Djafari-Rouhani 2009 Appl. Phys. Lett. **95** 124104.
[23] V. Laude, M. Wilm, S. Benchabane, A. Khelif 2005 Phys. Rev. E **71** 036607.
[24] Z. Cheng, Z. Shi 2014 Constr. Build. Mater. **50** 257.
[25] G. Futhazar, W.J. Parnell, A.N. Norris 2015 J. Sound Vib. **356** 1.
[26] J. Li, L. Fok, X. Yin, G. Bartal, X. Zhang 2009 Nat. Mater. **8** 931.
[27] T.-T. Wu, Z.-G. Huang, S. Lin 2004 Phys. Rev. B **69** 094301.
[28] X. Zhang, Z. Liu 2004 Appl. Phys. Lett. **85** 341.
[29] S. Benchabane, A. Khelif, J.Y. Rauch, L. Robert, V. Laude 2006 Phys. Rev. E **73** 065601.
[30] Y. Wang, F. Li, W. Huang, Y. Wang 2008 J. Mech. Phys. Solids **56** 1578.
[31] Y. Wu, K. Yu, X. Li, H. Zhou 2016 J. Phys. D: Appl. Phys. **49** 025502.
[32] A. A. Cannarozzi, F. Ubertini 2001 Int. J. Solids Struct. **38** 717.
[33] A. Nayfeh, S. Nemat-Nasscr 1971 Acta Mech. **12** 53.
[34] V.K. Agarwal 1978 J. Elasticity **8** 171.
[35] A. H. Nayfeh, Wave propagation in layered anisotropic media: with application to 29 composites, Elsevier, The Netherlands, 1995.
[36] S. Asiri, A. Baz, D. Pines 2006 Smart Mater. Struct. **15** 1707.
[37] J. Yu, B. Wu, C. He 2011 Acta Mech. Solida Sin. **24** 135.